# Berthil Cepstrum: a Novel Vibration Analysis Method based on Marginal Hilbert Spectrum Applied to Artificial Motor Aging

**Harun Šiljak**

*International Burch University*

*Francuske revolucije bb, 71000 Sarajevo, Bosnia and Herzegovina*

*harun.siljak@ibu.edu.ba*

**Abdulhamit Subasi**

*College of Engineering, Effat University*

*Jeddah, 21478, Saudi Arabia*

*absubasi@effatuniversity.edu.sa*

*Abstract*

Motor age determination as a part of condition monitoring heavily employs vibration analysis. This study introduces a new method for such analysis, based on concepts of cepstrum and marginal Hilbert spectrum. This new method, named Berthil cepstrum may be applied in general signal processing, not only when vibration signals are concerned. Classical marginal Hilbert spectrum has also been applied to the artificial motor aging data with excellent results. Furthermore, a ranking of known spectrum-based methods for determination of motor age together with the new methods introduced in this study has been made based on SVM and RELIEF attribute ranking, showing quality of the new methods.

**NOMENCLATURE**

AMIF – Automutual Information Function

EMD – Empirical Mode Decomposition

$H^3$VD – Hilbert-Hurst-Higuchi Vibration Decomposition

HHT – Hilbert Huang Transform

HVD – Hilbert Vibration Decomposition

IMF – Intrinsic Mode Function

PSD – Power Spectral Density

SVM – Support Vector Machine

## 1. INTRODUCTION

Motor condition monitoring is an important engineering problem in both theory and practice, with various methods introduced so far.[1] Different parameters have been monitored in order to provide relevant input for the analysis.

One of the often used parameters is the motor vibration, while others may include stator current or sound. [2,3] The vibration signals collected from the motor at various places are usually employed to detect bearing and shaft faults through analysis in both time and frequency domain, often decomposed using a method like wavelet transform. [2]

A particular problem in condition monitoring and predictive maintenance is determining the motor age and remaining useful life (RUL).[4] This task is difficult due to different damage the motor can encounter. A good way of facing most of the possible damage scenarios is artificial motor aging experiment based on IEEE Std-117.[5] The motor can be monitored at different stages of its aging, starting from a new motor and ending with a broken, non-functioning one after suffering chemical, electrical and thermal stress.

This study falls in the category of spectral methods for vibration analysis. In course of wider research, multitude of other methods were tested and their results presented. [6,7,8] However, an open question before this work was which of those methods is the best with respect to this particular problem.

This paper aims at presenting a new method and showing ist superiority to its classical counterpart. While it is based on rather well-known marginal Hilbert spectrum which has previously been applied to bearing faults, Berthil cepstrum is a novel tool, presented here for the first time. [9,10] This new tool is supposed to work as a general signal processing technique, applicable outside the vibration analysis discipline as well.

Results of this work will be presented in the following manner. Second section introduces motor aging as the source of vibration data being analyzed by the novel method. In the third section, existing signal processing methods previously applied to artificial motor aging vibration data together with the novel Berthil cepstrum are introduced, as well as a scheme for ranking the methods by their effectiveness. Fourth section gives an overview of application results which are thoroughly discussed in section five. Finally, conclusions are given in the last section.

## 2. MOTOR AGING

### *2.1. The Fundamentals*

The focus of our consideration is on motor bearing faults. It is important to note that bearings with a non-conductive grease film support the rotor in the motor construction, the film evenly distributing itself on the rolling surfaces. If the rotor voltage increases over a certain threshold, the grease film may break down and discharge mode current running through the bearing may cause serious damage to it. Of course, there is a conducting mode in which current flows through the bearing without damage in its regular working process, but in the case of high, film-breaking voltage, the discharge is made through arching. The damage mechanism can be in a form of pitting, fluting, or potentially other types of damage. Name fluting originates from the groove-like “flutes” emerging from repeated pits in the bearing, caused by the repeated discharge.[11]

### *2.2. The Experiment*

This particular electrical discharge from the shaft to the bearing in order to cause bearing damage was induced in an experimental setup shown in Fig. 1.[12]

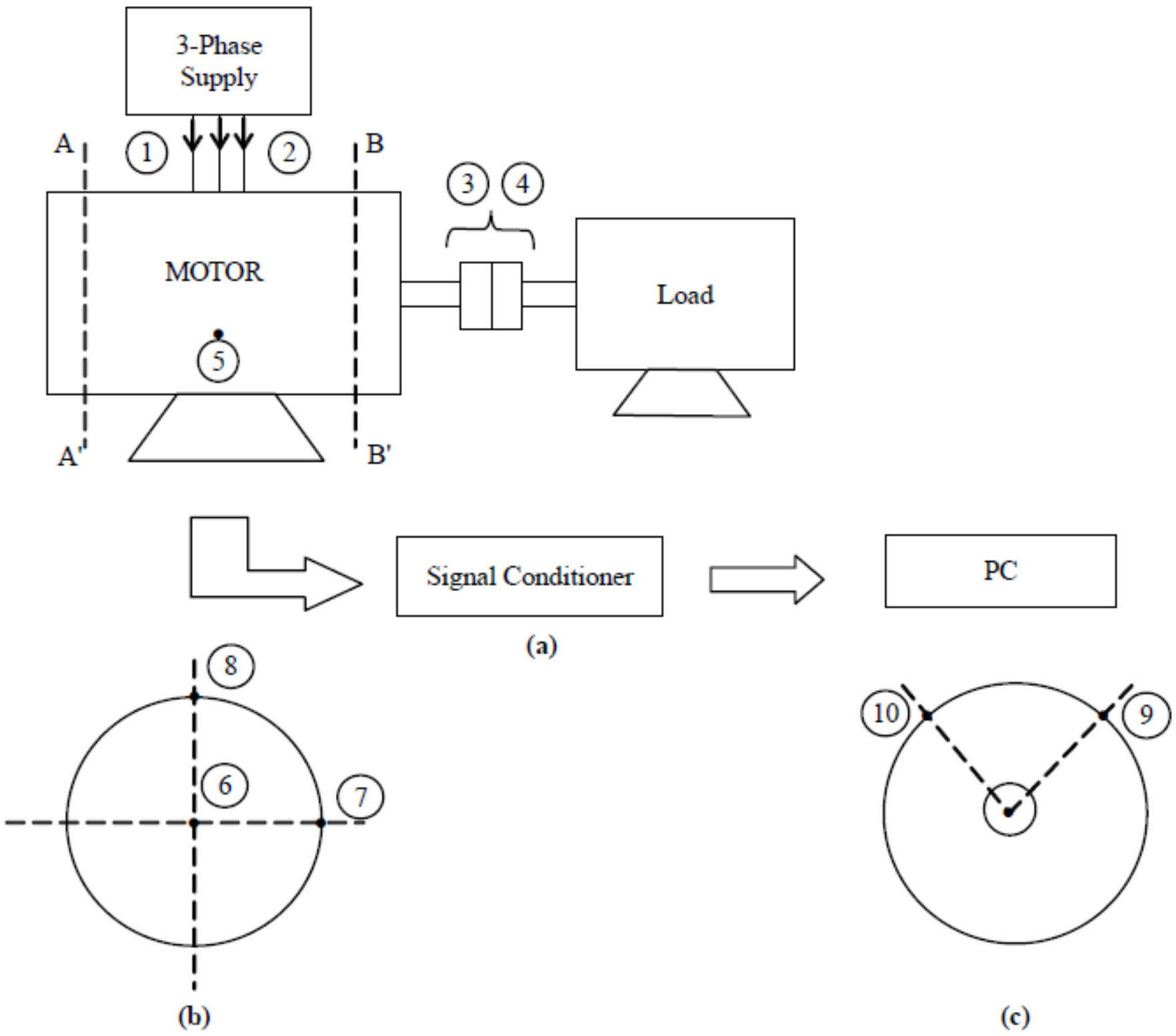

Figure 1. The experiment setup: (a) configuration, (b) and (c) cross-section AA' and BB' respectively. Numbers 1-4 denote electrical and mechanical measurement sensors, 6-10 are accelerometers.[11]

The fluting run of 30 minutes consisted of the motor running with no load and with an external 27 A, 30V AC current to the shaft. In each cycle's end, the motor subject to aging was run on full load in order to measure its performance, recording data at a sampling frequency of 12kHz. motor the test motor was put on a motor performance test platform. From the experimental set-up, high frequency data with a sampling frequency of 12kHz was collected. In this particular case, we are working with eight time series 10 seconds long (120,000 samples) representing motor condition from healthy case (state 0) to the last working case (state 7).

## 3. VIBRATION SIGNAL PROCESSING AND CLASSIFICATION

In this section, an overview of some simple methods applied earlier to the data from described experiment is given. It is followed by a presentation of the classical marginal Hilbert spectrum calculation and the novel Berthil cepstrum.

### *3.1. Fourier spectrum-based methods*

The methods based on Fourier spectrum are usually surrogate-invariable, meaning that they work for any data with the same amplitude spectrum, such as data produced by phase-randomizing surrogate generators.[13] While it is possible to use results of Fast Fourier Transform (FFT) directly, use of power spectral density (PSD) estimates based on autoregressive models is popular due to clear representation of characteristic features of PSD.[14] One of such methods is the Burg method, using Burg autoregressive model to estimate PSD.[15] Basic statistical analysis conducted on Burg PSD estimates may lead to useful indicators of motor age, namely minimum, mean and standard deviation have been used before. [6]

Another way to represent signal's spectrum is the use of cepstrum. [16] (Real) cepstrum is defined as the inverse Fourier transform of the Fourier transformed signal's (real) logarithm, i.e. $C(t)=F^{-1}\{\log(F(x(t)))\}$. Since it is technically a time-domain function, but representing the frequencies, it's abscissa is usually called *quefrency* (anagram of *frequency,* just like *cepstrum* is an anagram of *spectrum*). Previous work showed that the standard deviation of cepstra may also serve as an indicator of motor age. [6]

A very crude way of determining overall frequency of a signal is counting the number of zero crossings (of the signal having its non-zero mean subtracted). It is also considered an indicator of age. [6]

Finally, another property related to the Fourier spectrum is the Hurst Exponent.[17] Defined to be the measure of system's long-term memory, Hurst exponent takes values between 0 and 1, and in case of systems with long-term positive autocorrelation, it is larger than 0.5, while negative autocorrelation is indicated by the exponent lower than 0.5. Its relationship with the Fourier spectrum is through the spectrum slope, which is a linear function of the Hurst exponent. While it has been shown that Hurst exponent is not particularly suitable for age determination, it serves as a part of a method which has shown its efficiency in this task. [18]

This method, named Hilbert-Hurst-Higuchi Vibration Decomposition, relies on calculation of Hurst exponent (using the so-called Higuchi algorithm) for components of the original signal obtained through Hilbert Vibration Decomposition. [7,19]

Hilbert Vibration Decomposition, introduced by Feldman, aims at extracting the harmonics with highest energy from the signal.[20] This decomposition is originally designed to prepare the signal for Hilbert transform in order to find the instantaneous frequency of signal (actually, instantaneous frequencies of its components), as the original signal usually does not meet the demands of Hilbert transform applicability. [21]

In practice, this means that the signal is decomposed in multiple components (modes), and Hurst exponent is calculated for each one. It has been shown that all the components show monotonic trends and are therefore readily applicable for age determination.

### *3.2. Application of Hilbert Marginal Spectrum*

With the already presented Fourier spectrum based methods in mind, the Hilbert transform application goes along the lines of making analogous results. Since the conditions of Bedrosian and Nuttall theorems do not hold for general signal forms encountered in signal processing, certain decomposition is almost always required in order to apply the Hilbert Transform.[21]

One of such methods is the already presented HVD, while another one goes by the name Hilbert Huang Transform (HHT).[22, 23] Essence of this transform is in its first part, the so-called Empirical Mode Decomposition (EMD), which is an empirical method of decomposition of a signal into its Intrinsic Mode Functions (IMF), which are simple oscillatory modes, having same number of extrema and zero crossings and zero-mean envelopes.

Sifting is the process of obtaining IMFs, and it is defined in the following manner:

1. Finding all local minima and maxima of the data in question

2. Connecting all maxima with one cubic spline (upper envelope) and all minima with another (lower envelope)

3. Subtracting the mean of envelopes from the signal. If the result is not satisfying the IMF condition of having same number of extrema and zero crossings, this procedure is repeated until the condition is met. Once it is finally met, the obtained IMF is subtracted from the original data, and remainder serves as the new “original data”.

There are various stopping conditions for sifting. In this particular case, we have used the stopping criterion proposed by Rilling et al[23] where the fluctuations of the remainder in the most of the signal (both Rilling et al and us used 95%) are below a certain threshold, but allowing the remaining 5% of the remainder to have larger fluctuations.

It is not known beforehand how many IMF components are to be found within a signal with such a stopping condition: in case of our data, we were working on 16 to 19 IMFs.

Once these components are obtained, Hilbert transform can be applied to them in order to find the instantaneous frequency, based on the definition of so-called analytic signal, which produces a time-frequency relation. Since our Fourier approach was not a time-frequency, but a pure frequency analysis, this time-frequency result is not comparable to it. Hence, the results of Hilbert transform have to be represented in a pure frequency variant, resembling that of Fourier spectrum. For that purpose, Hilbert marginal spectrum is appropriate, as it is an integral in time of the classical Hilbert spectrum[24]:

$$h(w)=\int H(w,t)\,dt$$

Here, $T$ denotes the full data length, while the classical Hilbert spectrum $H$ is defined as

$$H(w,t)=\Re\sum a_i(t)e^{j\Phi_i(t)}$$

with

$$\Phi_i=\arctan\frac{H[c_i(t)]}{c_i(t)},a_i(t)=\sqrt{c_i^2(t)+H^2(c_i(t))}$$

In this calculation, $c_i$ denotes the $i$-th IMF, and H($c_i$($t$)) is its Hilbert transform

$$H[c_i(t)]=\frac{1}{\pi}\int\frac{c_i(\tau)}{t-\tau}d\tau$$

Of course, instead of IMFs, one could use the results of HVD as well, and the differences between these approaches have already been discussed in past.[25]

If we apply this Hilbert marginal spectrum calculation to data from artificial aging, and compare the spectrum obtained this way with Fourier spectrum, result is represented in Figure 2. One may suggest that peaks are more clearly seen in the Fourier spectrum, but the classification experiment will show the advantages in applying the Hilbert spectrum. At this point, it should be noted that Flandrin's EMD scripts for MATLAB are used to find IMFs and Hilbert spectrum, while Saa's marginal Hilbert spectrum script was used for the final result.[26,27]

In order to have a numerical attribute from the Hilbert marginal spectrum, spectrum centroid is calculated, using the formula

$$M=\frac{\sum f_k a_k}{\sum a_k}$$

where $a_k$ is the spectrum magnitude in $k$th frequency bin, and $f_k$ is the frequency of the bin. This value, calculated for each signal's spectrum is used as one of the attributes in age classification.

While the results from the previous subsection were obtained from our previous research, the results of Hilbert marginal spectrum application (in particular, using its centroid as a determinant of age) to artificial motor aging data are presented here for the first time.

### *3.3 Berthil Cepstrum*

As mentioned earlier, cepstrum is defined as $C(t)=F^{-1}\{\log(F(x(t)))\}$. If we replace the Fourier transform with marginal Hilbert spectrum, we get $B(t)=F^{-1}\{\log(h(x(t)))\}$, a transform we named Berthil cepstrum (an anagram of Hilbert, in the same spirit as cepstrum-spectrum). Note that, while the direct Fourier transform is replaced by Hilbert marginal spectrum, the inverse Fourier transform is not. The visual difference between the two cepstra on real data is shown in Figure 3.

(a)

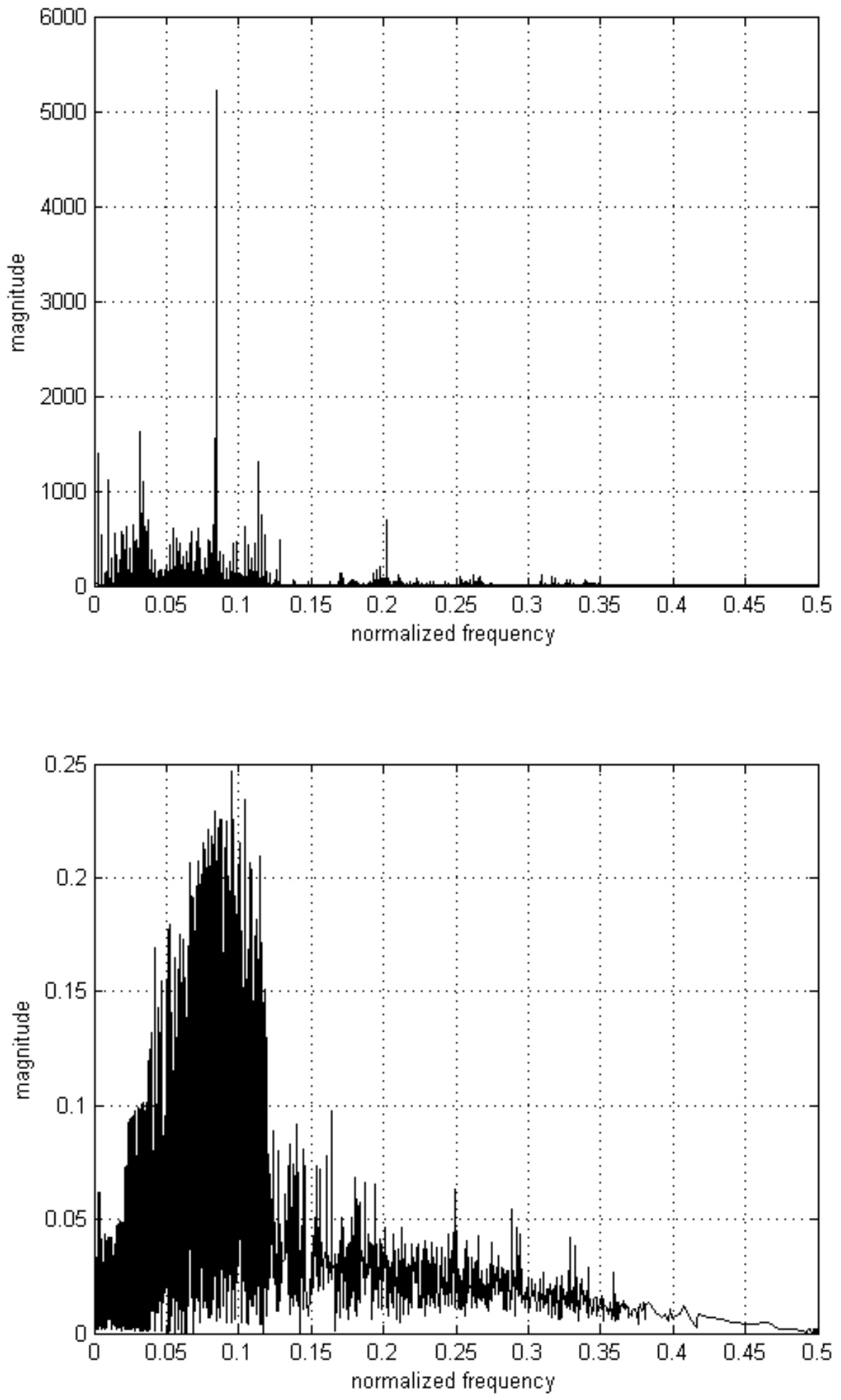

(b)

Figure 2. (a) Fast Fourier Transform of the healthy case vibration data (b) Marginal Hilbert Spectrum of a part of healthy case vibration data

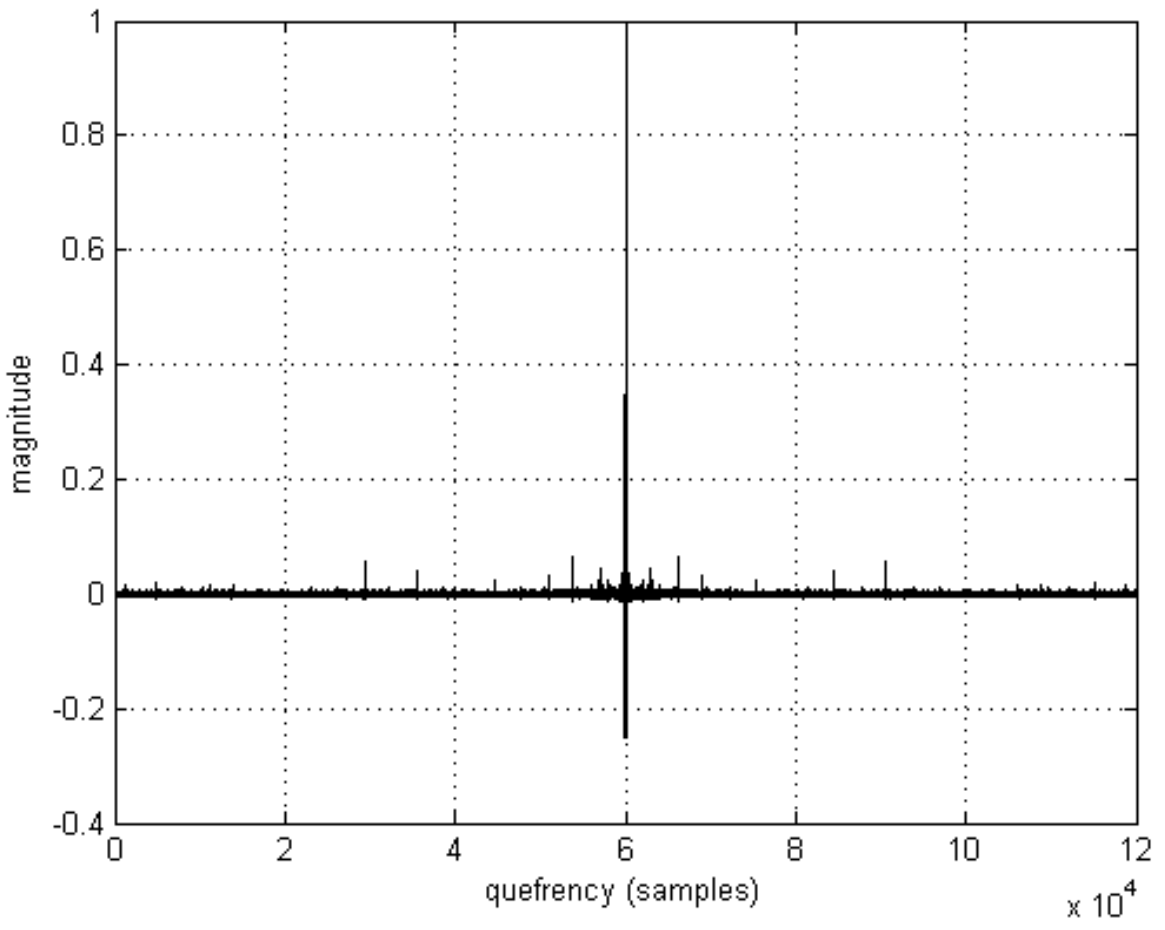

(a)

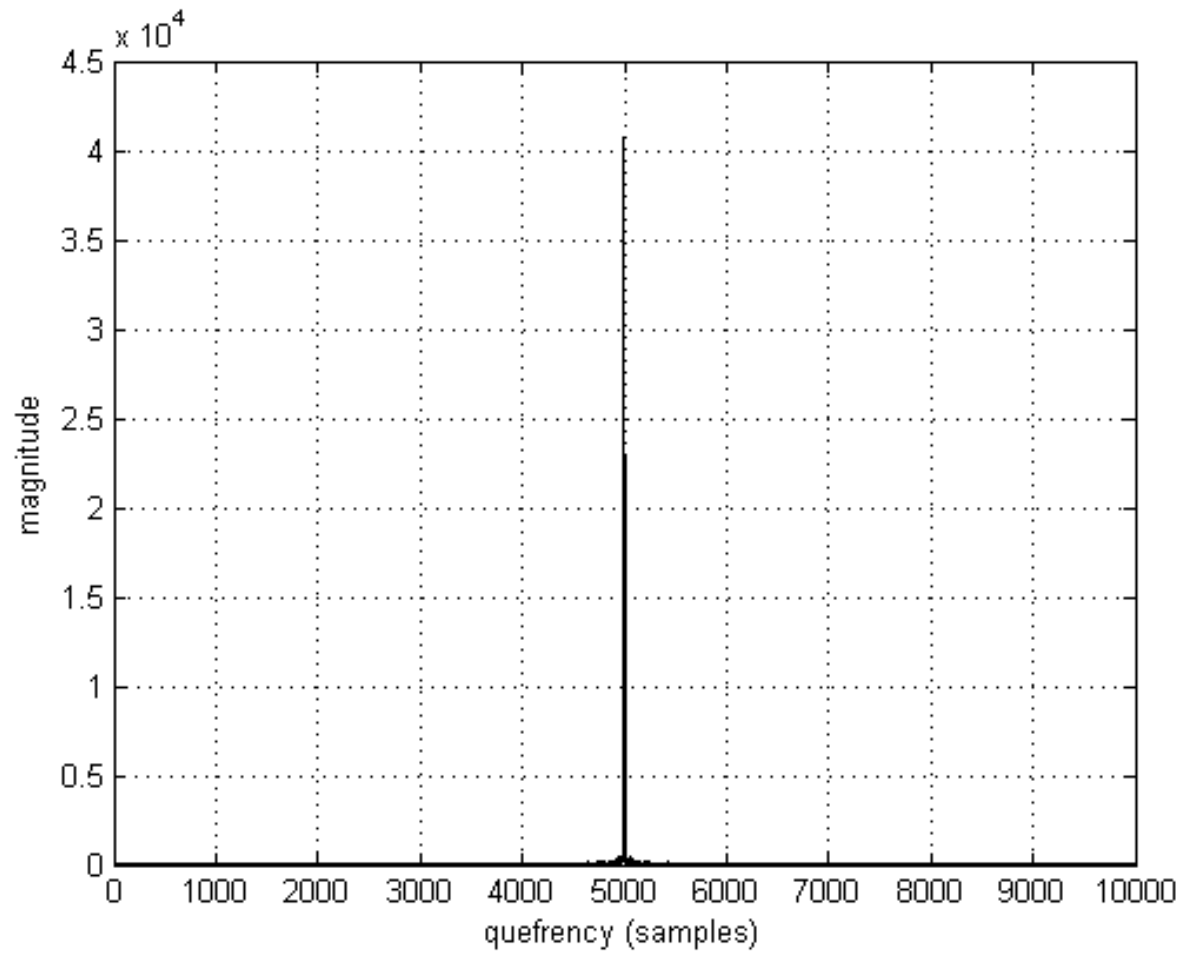

(b)

Figure 3. (a) Classical real cepstrum of the healthy case vibration data (b) Berthil cepstrum of the healthy case vibration data

The first important difference visible from Figure 3 is the existence of certain peaks in the classical spectrum which are not present in the Berthil spectrum. While these peaks could be used in detection of fault frequencies, results based on locating these peaks are bad, as reported in previous surveys.[28] Another difference between the two cepstra will be explained in the following section.

Justification for introducing this method in vibration analysis is given through the results and discussion to follow, in comparison with other proposed methods and their results.

Mathematical (i.e. theoretical) justification will surely be of interest for future research, and for the time being results remain purely empirical.

### *3.4. Mean and standard deviation of cepstra*

In this section, we will present a result on known artificial data which suggests applicability of Berthil cepstrum in classification.

The data used here are 1 second long signals of the format

$$w_d(t) = 2\sin(800\pi t) + \sin(120\pi t) + 0.5\sin(240\pi t) + 0.2\sin(500\pi t) + d\sin(6000\pi t) + \varepsilon(t)$$

where *d* is a variable taking values of 0.2, 0.5 and 1 (which is analogous to increase of power of vibration at higher frequencies observed in artificial aging [6,12]) and $\varepsilon(t)$ is the additive Gaussian white noise with an SNR of 5.

Both classical and Berthil cepstra were calculated for these three signals and low order statistics of cepstra (mean and standard deviation) were determined (Table 1).

Table 1. Low order statistics of cepstra normalized to 1-10 range

| Signal no/statistic | I | II | III |
|---:|---:|---:|---:|
| Cepstrum mean | 1.1290 | 1.5343 | 2.3185 |
| Berthil mean | 1.8301 | 1.9337 | 1.9966 |
| Cepstrum st. dev. | 2.9200 | 2.9220 | 2.9336 |
| Berthil st. dev. | 8.9944 | 8.6172 | 6.7853 |

The following section will show that the standard deviation is a good indicator of motor age. At this point, we see that Berthil cepstrum's standard deviation changes more (i.e. its own standard deviation is larger) than classical cepstrum's standard deviation, making it more suitable to serve as an attribute for motor age classification.

### *3.5. Attribute ranking*

Methods presented in previous subsections produce numerical attributes which are used to characterize the vibration signal in question. Although all of them show the same trends as it may be seen in the next section (Table 2), it cannot be said straightforwardly which of them is more suitable for classification/age determination than the others. Two suitable algorithms for ranking of numerical attributes are RELIEF feature selection and Support Vector Machine (SVM) feature selection.

RELIEF selection begins with scaling all $m$ features of $n$ instances of two known classes on the same range, [0,1]. In each of $p$ iterations of the algorithm, the procedure is repeated: feature vector of a random instance $\{x_i\}$ is taken, together with feature vectors of the closest instances from the same and the other class, $\{y_i\}$ and $\{z_i\}$, respectively. The weight vector $\{w_i\}$, which is initially zero-vector, is updated using the formula $w_i=w_{i-1}-(x_i-y_i)^2+(x_i-z_i)^2$. Finally, the weight vector is divided by the number of iterations and this scaled sequence is taken as relevance vector. [29]

In case of SVM feature selection, attributes are ranked using the square of elements in weight vector produced by the SVM classifier. [30]

**4. RESULTS**

In both Tables 2 and 3 presenting the results, the following abbreviations are used:

- AMIF: Automutual information function's first minimum
- $H^3$VDn (where $n$ is an integer between 1 and 7): Hurst exponent for Hilbert-Hurst-Higuchi Vibration Decomposition at $n$th component
- PSDMin: Minimum of Burg's PSD estimate
- PSDMean: Mean of Burg's PSD estimate
- PSDStd: Standard deviation of Burg's PSD estimate
- CepSTD: Standard deviation of real cepstrum
- BerSTD: Standard deviation of Berthil cepstrum
- ZC: number of zero-crossings relative to data length
- HilCEN: Hilbert marginal spectrum centroid

Table 2 summarizes the results of all listed methods from previous works, together with the new results for Hilbert marginal spectrum centroid and Berthil cepstrum standard deviation. All calculations were done in MATLAB.

To determine which of these numerical attributes is the most suitable one for motor age detection, the eight vibration signals from the experiments are divided in two classes: new (sets 0-3) and old motor (sets 4-7), and attribute ranking for this simple classification was conducted in Weka for all attributes in Table 2. Results in Table 3 are obtained for a slightly modified setup, in which the data set 4 is excluded for its already reported irregularity. However, SVM method gave the same results with or without data set 4 in the setup, while RELIEF differed. In Table 3, the attributes are listed in order from best to worst, and RELIEF relevance is given for reference.

Table 2. Numerical features of vibration signals subject to different spectrum-based features

| Data | AMIF | H3VD1 | H3VD2 | H3VD3 | H3VD4 | H3VD5 | H3VD6 | H3VD7 | PSDMin | PSDMean | PSDStd | CepSTD | BerSTD | ZC | HilCEN |
|---|---|---|---|---|---|---|---|---|---|---|---|---|---|---|---|
| 0 | 4 | 0.187 | 0.306 | 0.479 | 0.4783 | 0.5408 | 0.5529 | 0.5067 | -54.7 | -36.1 | 11.87 | 0.0057 | 419.92 | 0.1604 | 0.0589 |
| 1 | 4 | 0.1431 | 0.3512 | 0.4202 | 0.3956 | 0.337 | 0.3778 | 0.3234 | -44.93 | -28.06 | 8.857 | 0.0079 | 339.0935 | 0.2226 | 0.0804 |
| 2 | 4 | 0.0479 | 0.1284 | 0.1606 | 0.1065 | 0.2012 | 0.1249 | 0.1479 | -38.92 | -22.94 | 7.641 | 0.0091 | 307.9174 | 0.3242 | 0.1128 |
| 3 | 4 | 0.0328 | 0.0358 | 0.0509 | 0.0479 | 0.0539 | 0.072 | 0.0838 | -34.59 | -19.29 | 6.954 | 0.0095 | 225.2397 | 0.3663 | 0.1203 |
| 4 | 1 | 0.018 | 0.0237 | 0.0296 | 0.0397 | 0.0445 | 0.0508 | 0.0536 | -30.74 | -17.69 | 5.603 | 0.0099 | 290.126 | 0.4805 | 0.1618 |
| 5 | 1 | 0.0257 | 0.0309 | 0.0394 | 0.0438 | 0.053 | 0.0568 | 0.0612 | -31.68 | -17.23 | 6.339 | 0.0098 | 223.8033 | 0.4277 | 0.1524 |
| 6 | 1 | 0.0265 | 0.0254 | 0.0329 | 0.0409 | 0.0474 | 0.0523 | 0.058 | -29.85 | -15.79 | 6.116 | 0.01 | 200.4507 | 0.4473 | 0.1534 |
| 7 | 1 | 0.0154 | 0.0195 | 0.0236 | 0.0317 | 0.035 | 0.0424 | 0.048 | -26.57 | -13.09 | 6.086 | 0.0114 | 185.4076 | 0.4939 | 0.1697 |

Table 3. Attribute ranking based on SVM and RELIEF, with numerical values of RELIEF relevance for each method

| | 1 | 2 | 3 | 4 | 5 | 6 | 7 | 8 | 9 | 10 | 11 | 12 | 13 | 14 | 15 |
|---|---|---|---|---|---|---|---|---|---|---|---|---|---|---|---|
| SVM | AMIF | HilCEN | ZC | PSDMin | PSDMean | PSDStd | BerSTD | H3VD2 | H3VD3 | H3VD7 | CepSTD | H3VD5 | H3VD1 | H3VD6 | H3VD4 |
| RELIEF | AMIF | HilCEN | ZC | PSDMin | BerSTD | H3VD2 | H3VD3 | PSDMean | PSDStd | H3VD5 | H3VD7 | H3VD4 | H3VD1 | H3VD6 | CepSTD |
| Relevance | 1 | 0.359 | 0.301 | 0.218 | 0.214 | 0.21 | 0.207 | 0.207 | 0.183 | 0.16 | 0.149 | 0.144 | 0.139 | 0.129 | 0.122 |

For the graphical representation in Figure 4, all values from Table 2 were rescaled to the range [0,1] and grouped according to previous papers in which they appeared.

## 5. DISCUSSION

The fact that Marginal Hilbert Spectrum's centroid moves monotonically, with exception of data set 4, as well as the analogous behaviour of Berthil cepstrum standard deviation do not surprise, as such behaviour is observed in case of Fourier based approaches as well (as presented in Table 2).

Nevertheless, Berthil cepstrum's standard deviation is decreasing monotonically, while the (Fourier) cepstrum's standard deviation is increasing, which is indicating a qualitative difference in what these two representations of a signal really depict.

The comparison of Hilbert spectrum-based methods with earlier approaches made through attribute ranking as presented in Table 3 show the real potential of the methods presented here. Ignoring the AMIF as an ideal indicator in case of “old-new” classification, we see that Marginal Hilbert Spectrum's centroid is the best indicator, and Berthil Standard Deviation coming as fourth and sixth in case of RELIEF and SVM ranking, respectively. It is important to note that Berthil Standard Deviation appears to be better than its Fourier counterpart,

Cepstrum standard deviation, as well as $H^3VD$.

## 6. CONCLUSIONS

This study presented two methods for age determination in artificially aged induction motors. While the Hilbert marginal spectrum is not novel per se, its application for this data, shown to outperform the other methods by the feature selection algorithms is an important result. On the other hand, the Berthil cepstrum is a novel spectrum transform which may find its application in general signal processing. Results shown here imply its applicability in motor age determination and/or fault detection. Curious fact of opposite trending in case of classical cepstrum still remains an open question for analysis and discussion.

Future work may include further theoretical analysis of Berthil cepstrum, differences with respect to the classical cepstrum, possible modifications and potential areas of application. Another possible direction for future research is studying marginal Hilbert spectrum with some of the IMFs from the original signal left out, recognising the important modes showing the difference.

## ACKNOWLEDGEMENTS

The authors express their deepest gratitude to Prof. B.R. Upadhyaya and his team at the University of Tennessee, Nuclear Engineering Dept. for allowing use of the experimental data used here.

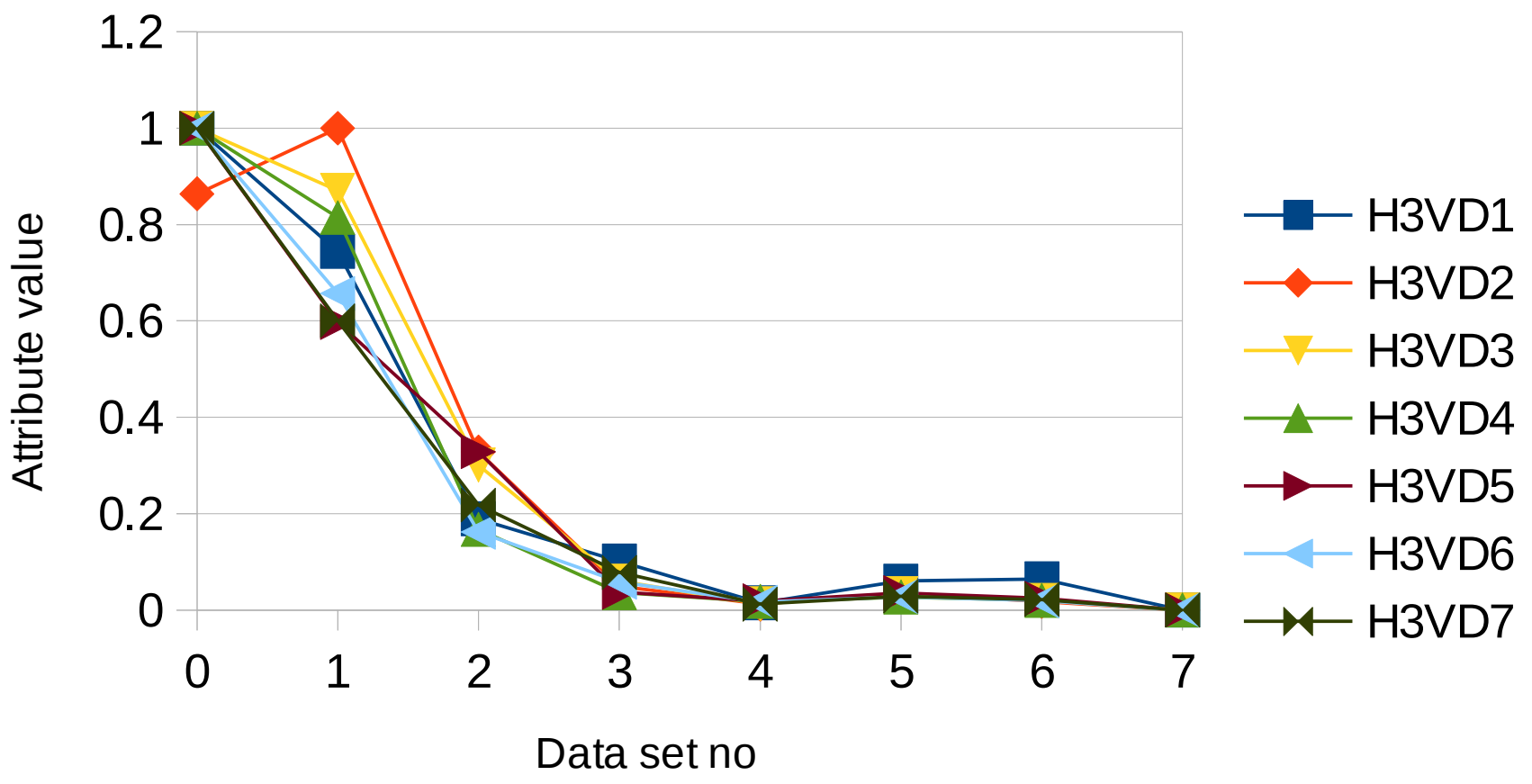

(a)

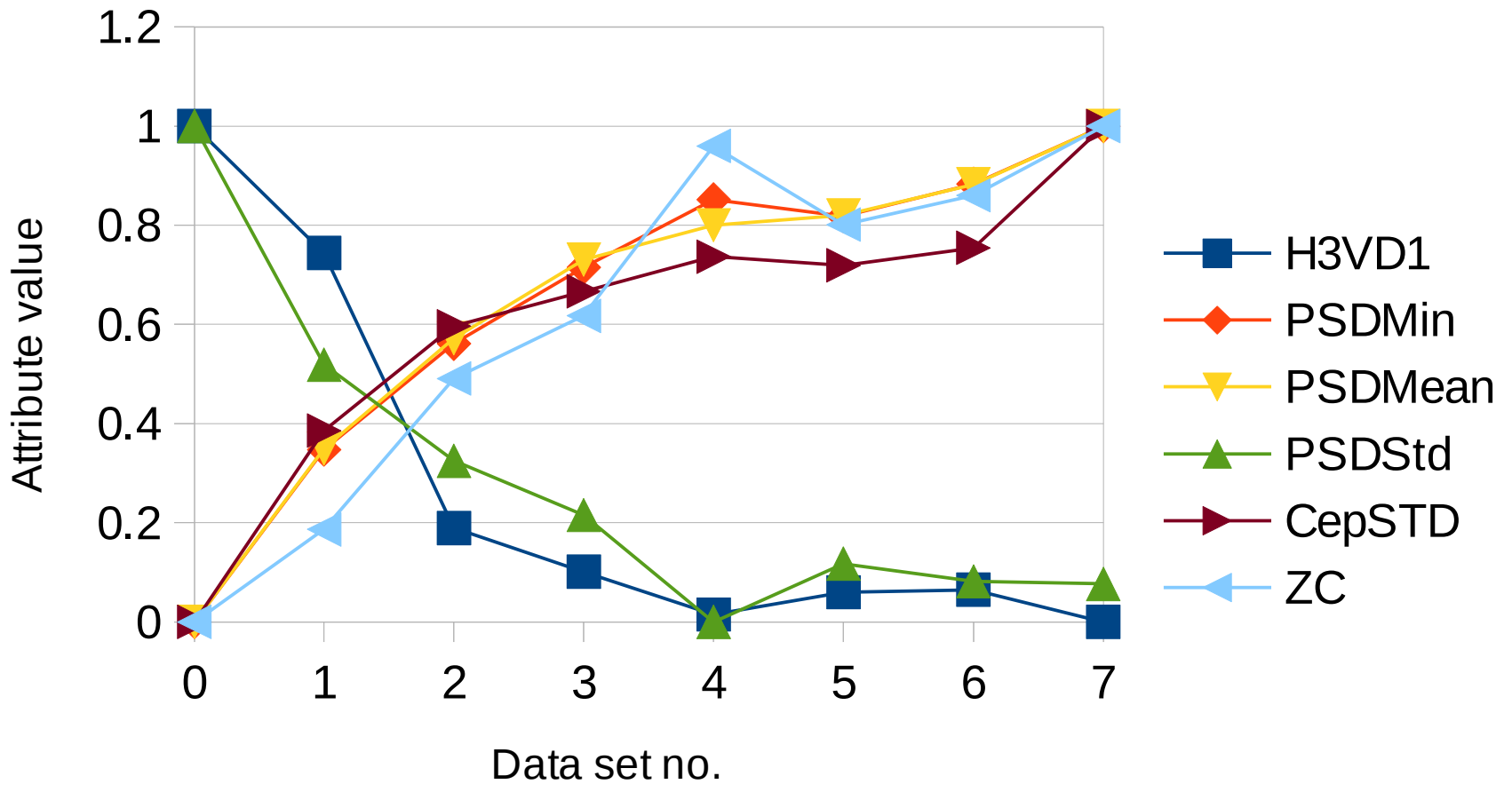

(b)

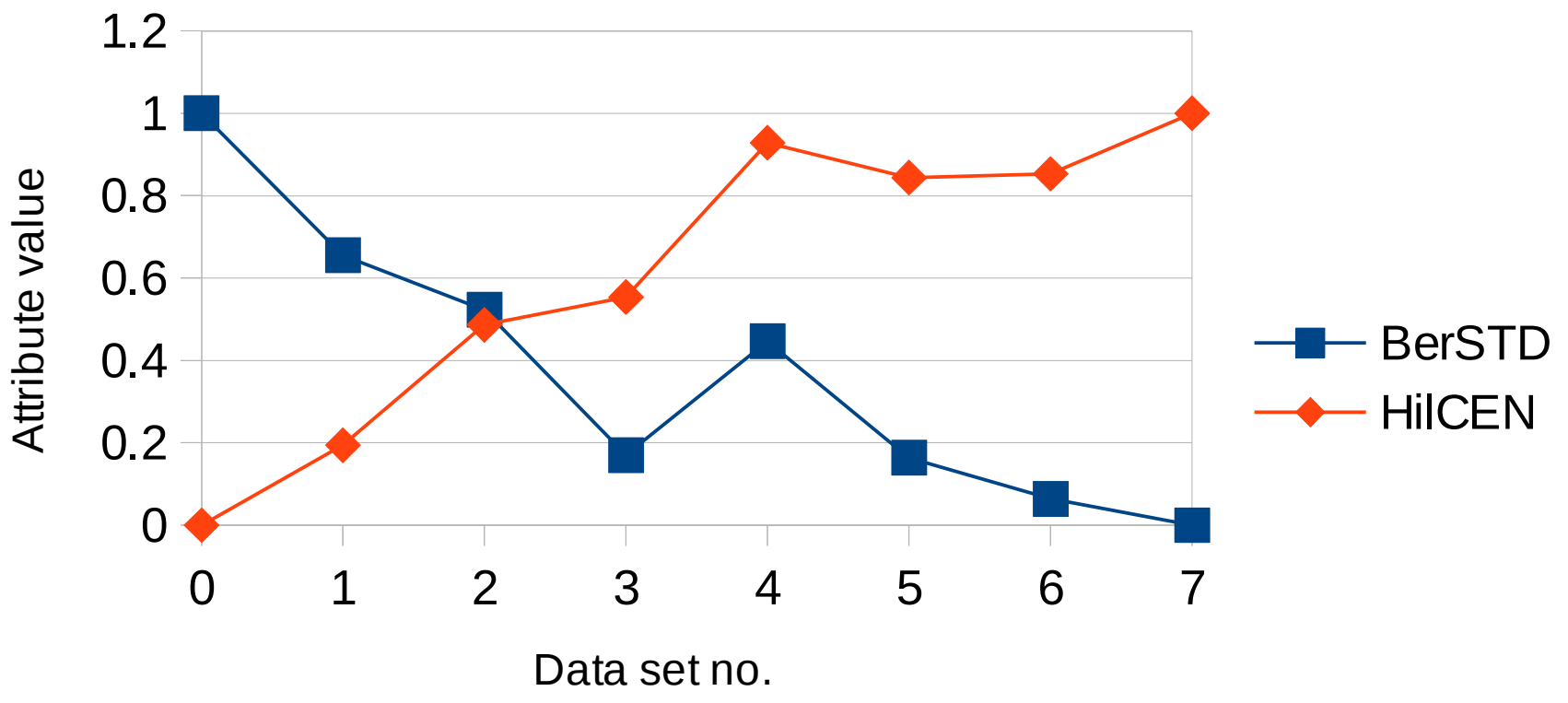

(c)

Figure 4. (a) $H^3VD$ numerical indicators (b) PSD, Cepstrum and Zero Crossing statistics (c) new Hilbert marginal spectrum based method indicators